# Dynamics of domain walls motion driven by spin-orbit torque in antiferromagnets

Luis Sanchez-Tejerina[1], Vito Puliafito[2], Pedram Khalili Amiri,[3] Mario Carpentieri[1], Giovanni Finocchio[4]

*[1] Dipartimento di Ingegneria Elettrica e dell'Informazione, Politecnico di Bari, Via Orabona 4, 70125 Bari, Italy*

*[2] Dipartimento di Ingegneria, Università di Messina, C.da Di Dio s/n, 98166 Messina, Italy*

*[3] Department of Electrical and Computer Engineering, Northwestern University, Evanston, IL, USA*

*[4] Department of Mathematical and Computer Sciences, Physical Sciences and Earth Sciences, University of Messina, Viale F. Stagno d'Alcontres 31, 98166 Messina, Italy*

***Abstract.*** The excitation of ultrafast dynamics in antiferromagnetic materials is an appealing feature for the realization of novel spintronic devices. Several experiments have shown that static and dynamic behaviors of the antiferromagnetic order are strictly related to the stabilization of multidomain states and the manipulation of their domain walls (DWs). Hence, a full micromagnetic framework should be used as a comprehensive theoretical tool for a quantitative understanding of those experimental findings. This model is used to perform numerical experiments to study the antiferromagnetic DW motion driven by the spin-orbit torque. We have derived simplified expressions for the DW width and velocity that exhibit a very good agreement with the numerical calculations in a wide range of parameters. Additionally, we have found that a mechanism limiting the maximum applicable current in an antiferromagnetic racetrack memory is the continuous domain nucleation from its edges, which is qualitatively different from what observed in the ferromagnetic case.

# I Introduction:

The nucleation and manipulation of ferromagnetic (FM) domain walls (DWs) have attracted a lot of attention in recent years due to the promising results for the development of spintronic devices such as racetrack memories,[1,2] memristors,[3,4,5] and sensors.[6] Nevertheless, the FM DW velocity, which is a key metric to evaluate the performance of those devices, driven by an external field drops beyond a certain field threshold (Walker breakdown),[7] while it saturates when an electric current is used as a driving force.[2,8] Recent experiments in synthetic antiferromagnets have demonstrated that the DW velocity can be as large as 750 m/s [9] and does not saturate within the applicable current ranges.[10] Ferrimagnetic DWs can also reach high velocities at the angular momentum compensation point as well.[11,12] In addition, it has been predicted that the velocity of DWs in antiferromagnets (AFM) should reach tens of km/s and it is limited by the group velocity of spin waves.[13,14,15] Here, we will focus on this latter category of materials due to their intriguing properties (absence of stray fields and low magnetic susceptibilities)[13,15,16,17,18,19] and potential importance either from a technological point of view, design of high-speed devices and better scaling in storage devices, and from a fundamental point of view to study the statics and the dynamics of multidomain states. Out of equilibrium, the antiferromagnetic order exhibits relaxation processes at ps time scale.[20,21,22] This THz dynamics makes those materials also appealing for the development of ultrafast spintronic devices.[19] In particular, on the path towards antiferromagnetic spintronics, antiferromagnetic domains can play the same role as the FM ones being the information carriers. The writing process can be achieved employing laser pulses[23] or spin-orbit-torques (SOT),[24,25] the manipulation by using alternating magnetic fields[13,23] or SOT, and the detection can be performed using one of the readout mechanism already observed experimentally such as tunneling anisotropic magnetoresistance (TAMR), anisotropic magnetoresistance (AMR), or spin Hall magnetoresistance (SMR).[26,27,28]

From a numerical point of view, antiferromagnetic dynamics can be described by atomistic micromagnetic models, [29] or at mesoscopic scale by a continuous micromagnetic framework that has proven to be very powerful for its ability to reproduce experimental observations in FM materials.[30] These models are based on the numerical solution of two Landau-Lifshitz-Gilbert equations, each of them describing one of the two sublattices of the antiferromagnet, strongly coupled through the exchange interactions. However, in the continuous formulation derived from atomistic models the exchange interactions are characterized by homogeneous, inter- and intra-lattice inhomogeneous terms at least.[31] Here, we perform an ideal numerical experiment to study the role of each of those exchange terms in the DW stability and dynamics.[22] In particular, we find that the homogeneous interlattice exchange does not affect the DW velocity and its role is limited to the stabilization of the antiferromagnetic order. On the other hand, the DW velocity as a function of either interlattice or intralattice inhomogeneous exchange field follows a square root dependence. We have derived simplified expressions for both DW size and velocity exhibiting a good agreement with numerical calculations that can be used for a fast exploration of DW statics and dynamics in a large space of material parameters. We also have found that the mechanism limiting the maximum applicable current is the nucleation of domains from the edges originating by boundary conditions imposed by the Dzyaloshinskii-Moriya interaction. In general, the continuous micromagnetic framework should be used for the qualitative understanding of recent switching experiments on antiferromagnetic devices, tens of microns in size, involving multiple domain states and memristive behavior.[3,4,5] The paper is organized as follows. In Section II the micromagnetic framework is described. Section III discusses the steps to derive the one-dimensional formulation. Results and conclusions are discussed in Section IV and Section V, respectively.

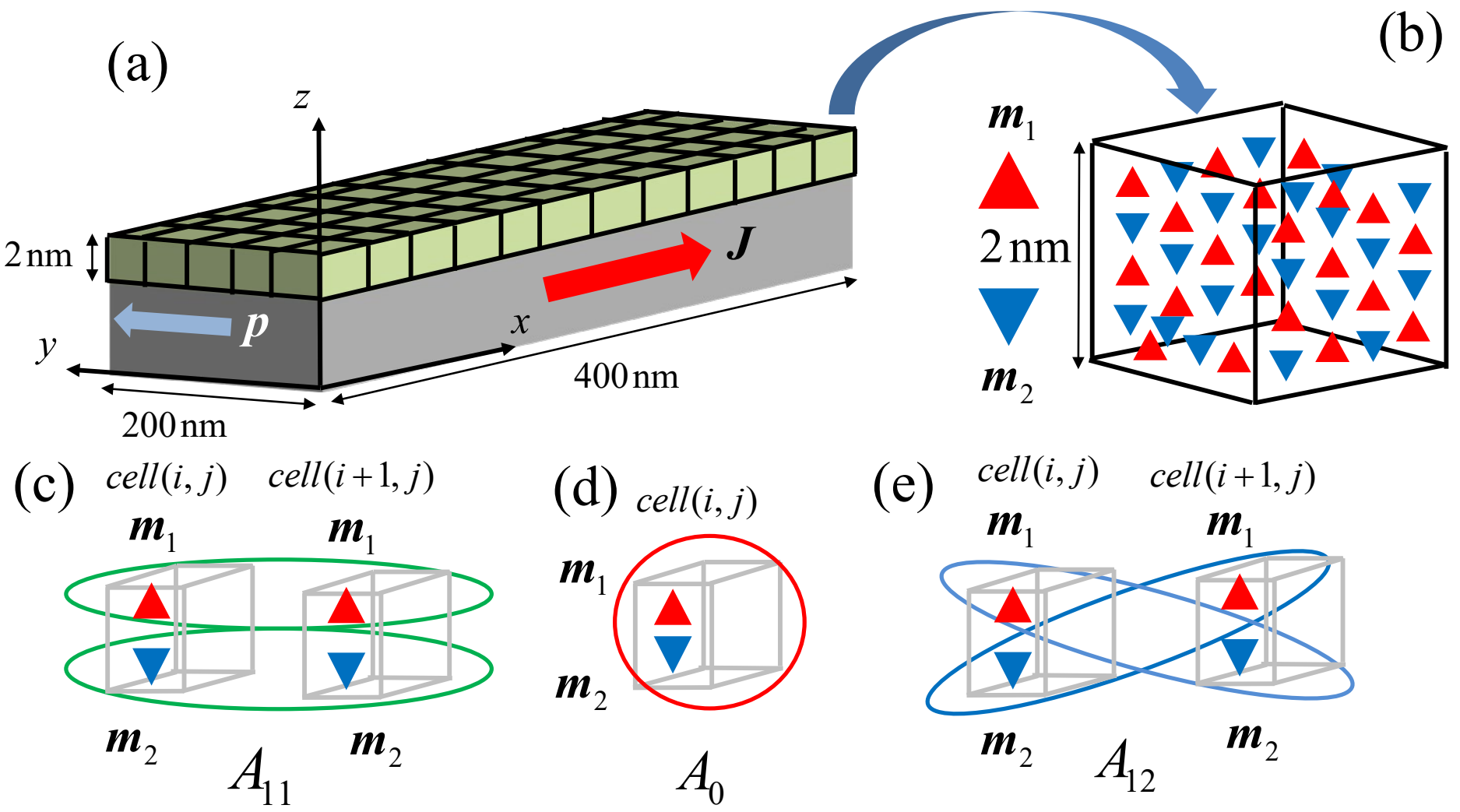

FIG. 1. (a) A schematic of the device under investigation characterized by antiferromagnetic material / heavy metal bilayer, with the indication of the Cartesian coordinate reference system and the device dimensions. The panel also includes the directions of the current density, $\boldsymbol{J}$, and the spin polarization, $\boldsymbol{p}$, and an example of the cubic discretization mesh (finite difference scheme) with a $2\,\text{nm}$ side used in this work to study the AFM. (b). Example of one computational cell with the magnetization vectors of the two sublattices $\boldsymbol{m}_1$ and $\boldsymbol{m}_2$. (c)-(e) Description of the three different exchange interactions included in this study, (c) inhomogeneous intralattice $A_{11} = A_{22}$, (d) homogeneous interlattice $A_0$ acting on the same computational cell, and (e) inhomogeneous interlattice $A_{12} = A_{21}$ both acting on the neighbors. Here, we consider the 6-neighbors for the computation of the inhomogeneous exchange terms indicated in (c) and (e).

## II Micromagnetic Model

*Device description.* Figure 1 shows a schematic of the system under investigation. It is a thin slab of an insulating AFM with perpendicular anisotropy, having lateral dimensions of $400 \times 200 \times 2\ \text{nm}^3$, on top of a heavy metal (HM) (e.g. Pt, Ta). A Cartesian coordinate system is introduced (see Fig. 1(a)) with the *z*-axis being the out-of-plane direction, while the *x* and *y*-axes are related to the length and the width of the device, respectively. The electric current is applied along the $x$-direction and flows in the HM layer, because of the SOT,[32,33,34] a spin density along the $y$-direction at the interface HM/AFM is accumulated.

*Model description.* Within the micromagnetic approach the AFM order is described by means of the magnetizations of two sublattices ($\boldsymbol{m}_1$ and $\boldsymbol{m}_2$) strongly antiferromagnetic coupled by the exchange interaction. We consider a finite difference discretization scheme (see Fig. 1) with the value of $\boldsymbol{m}_1$ and $\boldsymbol{m}_2$ reflecting the average magnetization of the spins within the same discretization cell. The AFM dynamics driven by the current can be described by the following LLG-Slonczewski equations,[35,36,37]

$$\begin{cases} \dfrac{d\boldsymbol{m}_1}{dt} = -\gamma_0 \boldsymbol{m}_1 \times \boldsymbol{H}_{eff,1} + \alpha \boldsymbol{m}_1 \times \dfrac{d\boldsymbol{m}_1}{dt} + \boldsymbol{\tau}_{SH,1} \\ \dfrac{d\boldsymbol{m}_2}{dt} = -\gamma_0 \boldsymbol{m}_2 \times \boldsymbol{H}_{eff,2} + \alpha \boldsymbol{m}_2 \times \dfrac{d\boldsymbol{m}_2}{dt} + \boldsymbol{\tau}_{SH,2} \end{cases} \tag{1}$$

where $\gamma_0$ is the gyromagnetic ratio and $\alpha$ is the Gilbert damping parameter, while

$$\boldsymbol{\tau}_{SH,i} = -\gamma_0 H_{SH}\, \boldsymbol{m}_i \times \left(\boldsymbol{m}_i \times \boldsymbol{p}\right) \tag{2}$$

is the antidamping SOT mainly due to the spin-Hall effect originating from a current density $J$ flowing through the HM,[32,33,34] with the amplitude given by $H_{SH} = \dfrac{\hbar \theta_{SH}}{2et\mu_0 M_S} J$ . In the last expression, $\hbar$ , $\theta_{SH}$ , $e<0$ , $t$ , $\mu_0$ are the reduced Planck's constant, the spin Hall angle, the electron charge, the AFM film thickness, and the vacuum permeability respectively. The saturation magnetization is equal in both sublattices $M_{S1} = M_{S2} = M_S$ . $\boldsymbol{p} = \boldsymbol{z} \times \boldsymbol{j}$ is the direction of the spin Hall polarization (see Fig. 1), $\boldsymbol{j}$ being the unit vector of the current density direction. Additionally, $\boldsymbol{H}_{eff,1}$ and $\boldsymbol{H}_{eff,2}$ are the effective fields for the first and second sublattice respectively.[38] They include the uniaxial anisotropy, the demagnetizing term, and the interfacial Dzyaloshinskii-Moriya interaction (iDMI) contribution.[8, 39] The exchange energy density can be written as

$$\varepsilon_{exch} = A_{11}\left(\nabla \boldsymbol{m}_1\right)^2 + A_{11}\left(\nabla \boldsymbol{m}_2\right)^2 - \frac{4A_0}{a^2}\boldsymbol{m}_1\boldsymbol{m}_2 + A_{12}\left(\nabla \boldsymbol{m}_1\right)\left(\nabla \boldsymbol{m}_2\right), \tag{3}$$

giving an exchange field with three contributions

$$\begin{aligned} \boldsymbol{H}_{1,exch} &= \frac{2A_{11}}{\mu_0 M_s}\nabla^2 \boldsymbol{m}_1 + \frac{4A_0}{a^2 \mu_0 M_s}\boldsymbol{m}_2 + \frac{A_{12}}{\mu_0 M_s}\nabla^2 \boldsymbol{m}_2, \\ \boldsymbol{H}_{2,exch} &= \frac{2A_{11}}{\mu_0 M_s}\nabla^2 \boldsymbol{m}_2 + \frac{4A_0}{a^2 \mu_0 M_s}\boldsymbol{m}_1 + \frac{A_{12}}{\mu_0 M_s}\nabla^2 \boldsymbol{m}_1. \end{aligned} \tag{4}$$

where $a$ is the lattice constant. In Eq. (3), the first term, $A_{11} > 0$, is the inhomogeneous intralattice contribution (Fig. 1(c)), the second one, $A_0 < 0$ , is the homogeneous interlattice (Fig. 1(d)), and the third, $A_{12} < 0$ , is the inhomogeneous interlattice contribution (Fig. 1(e)). The demagnetizing field is calculated by solving the magnetostatic problem[40] for the total magnetization $\boldsymbol{M}_{S1}+\boldsymbol{M}_{S2}$ where $\boldsymbol{M}_{Si}=M_S\boldsymbol{m}_i$. Our scheme is based on a field-based approach, so we compute directly the effective field rather than derive it from the energy density.[41] The antiferromagnetic material has been discretized into cubic cells with a side of 2 nm (Fig. 1 (b)). The following material parameters have been used:[22, 42, 11] lattice constant $a=0.35$ nm, saturation magnetization $M_S=0.4$ MA/m, uniaxial anisotropy constant $K_u=64\,\text{kJ/m}^3$, being $z$ its easy axis, spin Hall angle $\theta_{SH}=0.044$, Gilbert damping $\alpha=0.1$, and gyromagnetic ratio $\gamma_0=0.221\,\text{Mm/As}$. The expressions for the iDMI field are $H_{DMI,1}=-\frac{2D}{\mu_0 M_S}\left(\boldsymbol{u}_z(\nabla\cdot\boldsymbol{m}_1)-\nabla m_{1,z}\right)$, $H_{DMI,2}=-\frac{2D}{\mu_0 M_S}\left(\boldsymbol{u}_z(\nabla\cdot\boldsymbol{m}_2)-\nabla m_{2,z}\right)$, where the iDMI parameter $D=0.11\,\text{mJ/m}^2$ and $\boldsymbol{u}_z$ is the unit z-vector. In order to investigate the role of exchange fields in statics and dynamics, the exchange constants range from few pJ/m to few tens of pJ/m.

*Boundary conditions.* At the edges, the iDMI boundary conditions,[43] are determined by the fields $\boldsymbol{H}_{DMI,1S}=\frac{D}{\mu_0 M_S}\left(\boldsymbol{m}_1\times(\boldsymbol{n}\times\boldsymbol{u}_z)\right)$ and $\boldsymbol{H}_{DMI,2S}=\frac{D}{\mu_0 M_S}\left(\boldsymbol{m}_2\times(\boldsymbol{n}\times\boldsymbol{u}_z)\right)$ where $\boldsymbol{n}$ is the normal vector to the edge as in the case of the FM. In addition, it necessary to take into account the contribution from the interlayer exchange field, therefore, the boundary conditions for the *i*-th sublattice are given by the relation

$$2A_{11}\partial_n\boldsymbol{m}_i+A_{12}\boldsymbol{m}_i\times\left(\partial_n\boldsymbol{m}_j\times\boldsymbol{m}_i\right)+D\boldsymbol{m}_i\times\left(\boldsymbol{n}\times\boldsymbol{u}_z\right)=\boldsymbol{0}, \tag{5}$$

where $\boldsymbol{n}$ is the unit vector perpendicular to the edge.[44] At the right edge, we have:

$$\begin{aligned}
m\left(N_x+1,c_y,c_z\right)_{i,x}&=m\left(N_x,c_y,c_z\right)_{i,x}-\xi\left(m\left(N_x,c_y,c_z\right)_{i,z}-\frac{A_{12}}{2A_{11}}m\left(N_x,c_y,c_z\right)_{j,z}\right)\\
m\left(N_x+1,c_y,c_z\right)_{i,y}&=m\left(N_x,c_y,c_z\right)_{i,y}\\
m\left(N_x+1,c_y,c_z\right)_{i,z}&=m\left(N_x,c_y,c_z\right)_{i,z}+\xi\left(m\left(N_x,c_y,c_z\right)_{i,x}-\frac{A_{12}}{2A_{11}}m\left(N_x,c_y,c_z\right)_{j,x}\right)
\end{aligned} \tag{6}$$

where the couple $i$ and $j$ can be the following values, $i,j=(1,2)$ or, $i,j=(2,1)$ and $\xi=\frac{D\Delta x}{2A_{11}\left(1-\left(A_{12}/2A_{11}\right)^2\right)}$. $N_x$ is the number of computational cells along the $x$ direction, $\Delta x$ the lateral cell size, and $c_y$, $c_z$ refer to an arbitrary cell along $y$ and $z$ directions. Similar expressions are also valid for the other edges. For the special case $A_{12}=-2A_{11}$, a the Eq. (6) becomes

$$
\begin{aligned}
m\left(N_x+1,c_y,c_z\right)_{i,x} &= m\left(N_x,c_y,c_z\right)_{i,x} - \frac{D}{4A_{11}} m\left(N_x,c_y,c_z\right)_{i,z} \\
m\left(N_x+1,c_y,c_z\right)_{i,y} &= m\left(N_x,c_y,c_z\right)_{i,y} \\
m\left(N_x+1,c_y,c_z\right)_{i,z} &= m\left(N_x,c_y,c_z\right)_{i,z} + \frac{D}{4A_{11}} m\left(N_x,c_y,c_z\right)_{i,x}
\end{aligned}
\quad , \qquad (7)
$$

which is essentially the boundary condition of the ferromagnetic case for an exchange two times larger.

*Domain wall stability.* All the simulations were performed considering an antiferromagnetic Nèel DW type as a ground state that are stabilized by the iDMI. The equilibrium configuration has been computed by solving the equations $\begin{cases} \boldsymbol{m}_1 \times \boldsymbol{h}_{eff,1} = 0 \\ \boldsymbol{m}_2 \times \boldsymbol{h}_{eff,2} = 0 \end{cases}$ with a residual of $10^{-9}$. Figure 2(a) shows a snapshot (the color is related to the out-of-plane component) of a typical ground state of the two sublattices.

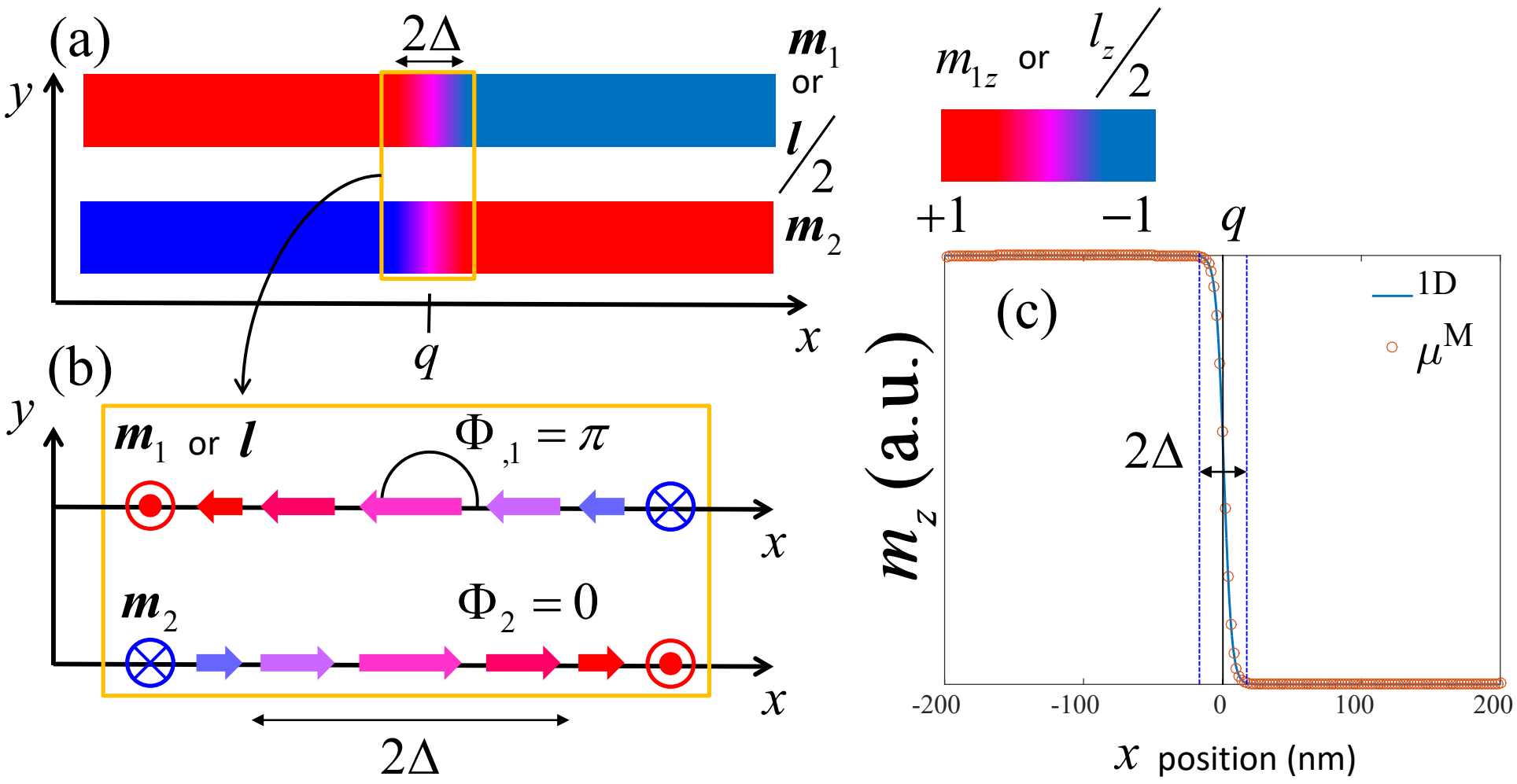

FIG. 2. (a) A snapshot of an antiferromagnetic Néel DW (the color is related to the out-of-plane component of the magnetization as indicated in the color bar), where its position, $q$, and its size, $\Delta$, are also indicated. (b) Definition of the Néel order parameter $\boldsymbol{l}$, the magnetization of the first sublattice $\boldsymbol{m}_1$ and the magnetization of the second sublattice $\boldsymbol{m}_2$ used in this work. $\Phi_1$ and $\Phi_2$ are the angles of $\boldsymbol{m}_1$ and $\boldsymbol{m}_2$ vectors with respect to the $x$-axis. (c) An example of the DW profile (z-component of the magnetization $m_{1,z}$) for the first sublattice as computed from micromagnetic simulations (empty circles) compared with the Walker ansatz (solid line) considering the parameters listed in Section II, and for $A_{11} = 15\,\text{pJ/m}$ and $A_{12} = 0\,\text{pJ/m}$. $q$ and $\Delta$ are also indicated for comparison with (a).

## III Analytical derivation of domain wall velocity and width

The derivation of the simplified expressions for the DW velocity and width is based on the 1-dimensional approximation (only the spatial dependence along the $x$-direction is considered) and assuming that the magnetization profile can be described by the Walker ansatz as reported in Eq. (8) (see Fig. 2(c) for a comparison with the micromagnetic profile).

$$\theta(t,x) = 2\arctan\left(\exp\left(Q\frac{x-q(t)}{\Delta(t)}\right)\right)$$
$$\Phi_i(t,x) = \Phi_i(t) \tag{8}$$

where $q$, $\Delta$, and $\Phi_1,\Phi_2$ are the DW position, width, and sublattice internal angles, as defined in Figs. 2(a) and (b). $Q=\pm1$ allows distinguishing between an up-down transition ($Q=1$) or a down-up transition ($Q=-1$).

*Simplified model.* Analytical models for the description of DW dynamics in AFM have been already derived. See for example Appendix A for the sigma model where the dynamics can be written in term of Néel order parameter $\boldsymbol{l}=\boldsymbol{m}_1-\boldsymbol{m}_2$ and the small magnetization $\boldsymbol{m}=\boldsymbol{m}_1+\boldsymbol{m}_2$. [13,14,35,42] Here, we develop a generalization of the previous models where: *(1)* the $\boldsymbol{l}$ -dependence of the homogeneous exchange and the $\boldsymbol{m}$ -dependence of the anisotropy are taken into account, *(2)* $A_{12}$ and $A_{11}$ are independent parameters, *(3)* $\Phi_1$ and $\Phi_2$ are free to evolve independently, and (4) the DW width is a dynamical variable $\Delta_1(t)=\Delta_2(t)=\Delta(t)$. Within these hypotheses, Eq. (1) in spherical coordinates reads

$$\begin{cases} \dot{\theta}_i = -\dfrac{1}{L\sin\theta_i}\dfrac{\delta u}{\delta\varphi_i} - \alpha\sin\theta_i\dot{\varphi}_i + \dfrac{1}{L}h_{SH}\cos\theta_i\sin\varphi_i \\ \sin\theta_i\dot{\varphi}_i = \dfrac{1}{L}\dfrac{\delta u}{\delta\theta_i} + \alpha\dot{\theta}_i + \dfrac{1}{L}h_{SH}\cos\varphi_i \end{cases} \qquad i=1,2 \tag{9}$$

being $L^{-1} = {}^{\gamma_0}\!/_{\mu_0 M_S}$, $h_{SH} = \dfrac{\hbar\theta_{SH}}{2et}J$. It is possible to compute a surface energy density from the integral of the energy density along $x$ and taking the Walker ansatz of Eq. (8), by making the hypothesis that $q_1=q_2=q$, $Q_1=-Q_2$. Within this assumption, the surface energy density is

$$\sigma = \frac{4A_{11}}{\Delta} - \frac{2A_{12}}{\Delta}\left(\frac{2-\cos(\Phi_1-\Phi_2)}{3}\right) - 2h_{exch}\Delta\left(\cos(\Phi_1-\Phi_2)+1\right) + \pi QD\left(\cos\Phi_1-\cos\Phi_2\right) +$$
$$+4K_u\Delta + \mu_0\Delta M_S{}^2N_x\left(\cos\Phi_1+\cos\Phi_2\right)^2 + \mu_0\Delta M_S{}^2N_y\left(\sin\Phi_1+\sin\Phi_2\right)^2 \tag{10}$$

where $h_{exch} = \dfrac{4A_0}{a^2}$ and $N_k$ ( $k=x,y,z$ ) are the demagnetizing factors associated with the DW corresponding to a prism having as size the strip width, the strip thickness and $2\Delta$. We wish to stress one more time that, differently from the sigma model, two different values for the intralattice and the interlattice inhomogeneous exchanges are considered. It is possible to link the surface energy density with the dynamic variables of the system through the LLG-Slonczewski Eq. (9), giving the relations between the variational derivatives and the partial derivatives of the surface energy density with respect to $q$, $\Delta$, $\Phi_1$ and $\Phi_2$. These relations lead to a set of differential equations describing the dynamics of the DW

$$\frac{\dot{q}}{\Delta}=\frac{\left[\alpha\frac{\pi}{2}h_{SH}\left(\cos\Phi_1-\cos\Phi_2\right)-\frac{\pi}{2}h_D\left(\sin\Phi_1+\sin\Phi_2\right)+2\left(h_{exch}'+h_{exch}\right)\sin\Phi_d+H_{d1}+H_{d2}\right]}{\left(1+\alpha^2\right)L_T}$$

$$\frac{\dot{\Delta}}{\Delta}=\frac{-12}{\pi^2\alpha L_T}\left[2K_u-\frac{2A_{11}}{\Delta^2}+h_{exch}'\left(2-\cos\left(\Phi_1-\Phi_2\right)\right)-h_{exch}\left(\cos\left(\Phi_1-\Phi_2\right)+1\right)+H_d'\right]$$

$$\alpha L\dot{\Phi}_1=H_{d1}+\frac{\pi}{2}h_D\sin\Phi_1-\left(h_{exch}'+h_{exch}\right)\sin\Phi_d+L\frac{\dot{q}}{\Delta}$$

$$\alpha L\dot{\Phi}_2=H_{d2}-\frac{\pi}{2}h_D\sin\Phi_2+\left(h_{exch}'+h_{exch}\right)\sin\Phi_d-L\frac{\dot{q}}{\Delta} \tag{11}$$

$$H_{di}=-\mu_0M_S^2\left(\left(N_x-N_y\right)\frac{\sin2\Phi_i}{2}+\left(N_x\sin\Phi_i\cos\Phi_j-N_y\sin\Phi_j\cos\Phi_i\right)\right)$$

$$H_d'=\frac{1}{2}\mu_0M_S^2\left(N_x\left(\cos\Phi_1+\cos\Phi_2\right)^2+N_y\left(\sin\Phi_1+\sin\Phi_2\right)^2\right) \tag{12}$$

$$i\neq j \qquad \Phi_d=\Phi_1-\Phi_2 \qquad L_T=2L$$

$h_{exch}'=A_{12}/3\Delta^2$ and $h_D=\pi QD/2\Delta$. The values of the two in-plane angles $\Phi_1$ and $\Phi_2$ are given by a trade-off between the torque exerted by the SHE, which tends to align the in-plane magnetization for each sublattice along the same direction, and the antiferromagnetic exchange energy that has a minimum for $\Phi_1=\pi+\Phi_2$. Additionally, once the values for $\Phi_1$ and $\Phi_2$ are reached, the DW width also acquires a stationary value

$$\Delta=\sqrt{\frac{2A_{11}-A_{12}\left(\frac{2-\cos\left(\Phi_1(J)-\Phi_2(J)\right)}{3}\right)}{2K_u-h_{exch}\left(\cos\left(\Phi_1(J)-\Phi_2(J)\right)+1\right)+H_d'(J)}}. \tag{13}$$

At these stationary conditions, taking the difference between the two last equations of the system Eq. (11), the expression for the DW velocity reads

$$\dot{q}=Q\Delta(J)\frac{\pi}{2}\frac{h_{SH}}{L_T\alpha}\left(\cos\Phi_1(J)-\cos\Phi_2(J)\right) \tag{14}$$

and, differently from Eq. (A6), the velocity depends on the stationary values of $\Phi_1$, $\Phi_2$, and $\Delta$ (all are a function of the applied current $J$). As compared to Eq. (A6), a first qualitative difference is that a saturation velocity is expected for large currents due to the transformation from Néel to Bloch ($\Phi_1 \to \pm\pi/2$ and $\Phi_2 \to \pm\pi/2$ depending on the current sign) domain wall similar to what is found in the ferromagnetic counterpart. Additionally, a decrement on the velocity with respect to the linear behavior is also expected due to the contraction of the DW width (note that $A_0 < 0$ so $-h_{exch} > 0$) However, for the parameters used in this work, a deviation from the linear behavior below the $0.2\%$ is expected for a homogeneous interlattice exchange of $A_0 = 0.5\,\text{pJ/m}$ and a current density $J = 1\,\text{TA/m}^2$ while is still below the $15\%$ for a current density of $J = 10\,\text{TA/m}^2$. Higher homogeneous interlattice exchanges would fit better with the linear behaviour and therefore these discrepancies are not easily observed experimentally.

The expression for the static DW width $\Delta$ is also derived from Eq. (13) taking into account that at equilibrium, no current is applied and $\Phi_1 = 0, \pi$, $\Phi_2 = \pi, 0$ so that the DW width reads

$$\Delta_e = \sqrt{\frac{2A_{11} - A_{12}}{2K_u}}. \tag{15}$$

This formula is a generalization of the expression for the DWs in FM[45] and it is a key result of this work. At equilibrium there are no misalignments between the two sublattices and, consequently, the static $\Delta_e$ does not depend on the homogeneous exchange. On the other hand, the two inhomogeneous exchange terms and the anisotropy have a key role in the determination of the DW size. The DW width $\Delta_e$ also does not depend on the iDMI parameter being its energy equal to $\sigma_{iDMI} = \pi Q D(\cos\Phi_1 - \cos\Phi_2)$. Full numerical simulations confirm this finding, showing that the $\Delta$ change is less than 1.5% while changing the iDMI parameter from $0.1\,\text{mJ/m}^2$ to $0.5\,\text{mJ/m}^2$.

## IV. Results and Discussions

*Static properties.* First of all, we have studied the static properties of the DW width by comparing calculations from micromagnetic simulations with Eq. (15). The results show a good agreement in a wide range of parameters. As an example, Fig. 3(a) and (b) summarize some of those comparisons. The square root dependence emerges when one of the two inhomogeneous exchange terms is zero (black line in Fig. 3(a)). The good agreement between numerical calculations and Eq. (15) confirms the lack of misalignments between both sublattices. Larger values of the inhomogeneous exchange contributions increase the DW width and, therefore, the minimum domain size.

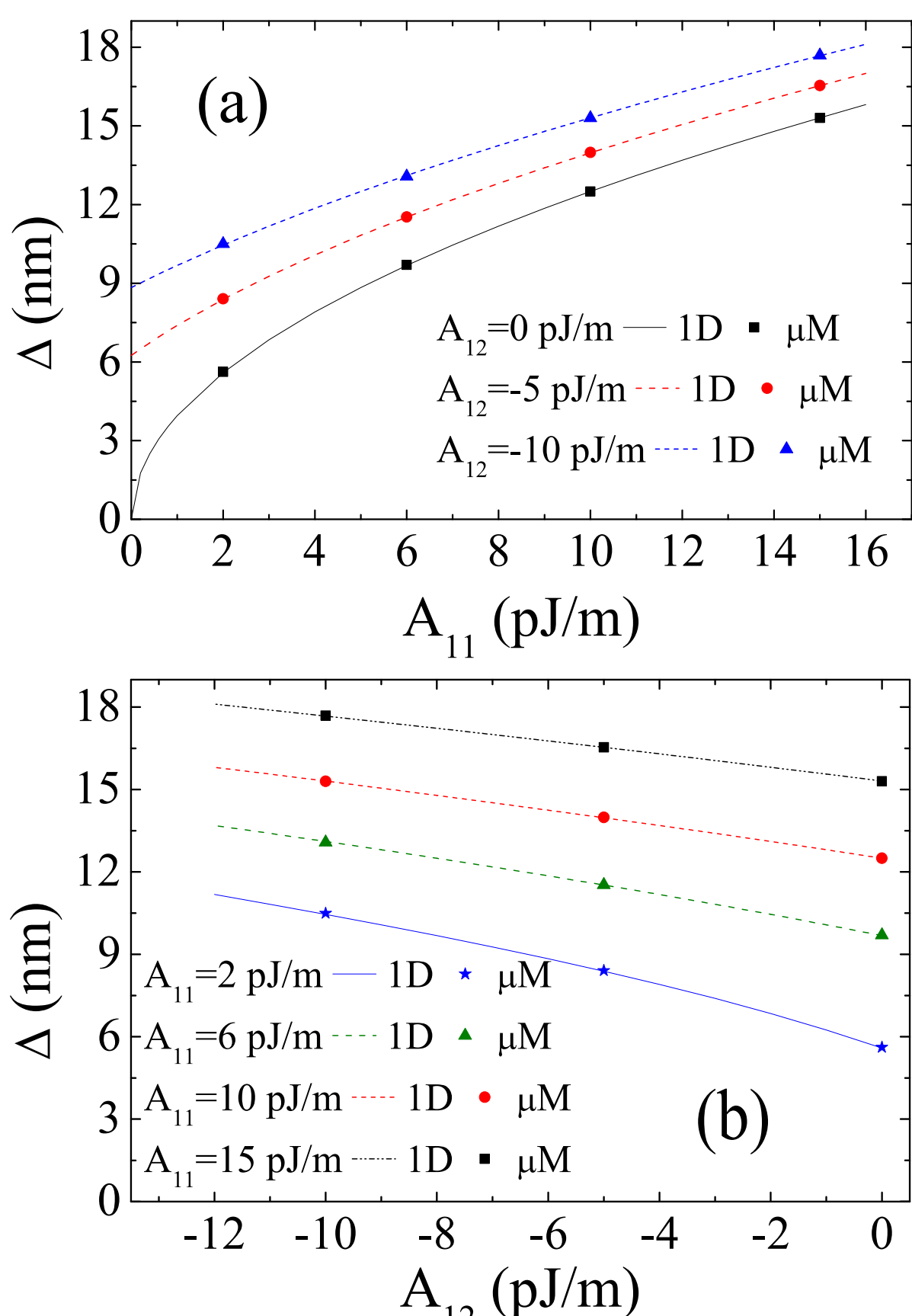

FIG. 3. DW width $\Delta$ as a function of (a) the intralattice inhomogeneous exchange, $A_{11}$, for three values of the interlattice inhomogeneous exchange ( $A_{12} = 0, -5, -10\,\mathrm{pJ/m}$ ), and (b) the interlattice inhomogeneous exchange, $A_{12}$, for four different values of the inhomogeneous intralattice exchange ( $A_{11} = 2, 6, 10, 15\,\mathrm{pJ/m}$ ). In both figures, the symbols stand for micromagnetic simulations and the solid lines are computed with Eq. (15).

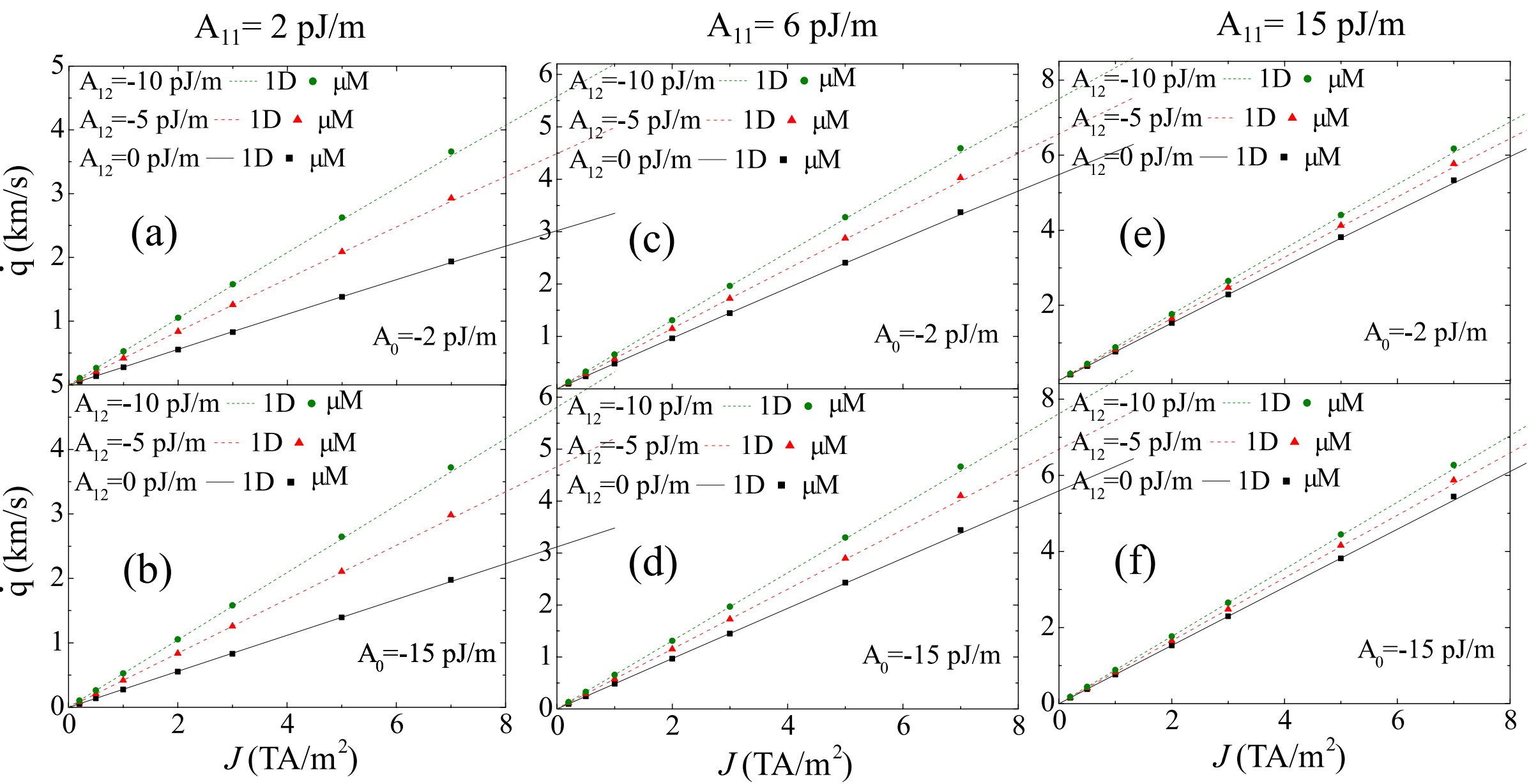

FIG. 4. DW velocity $\dot{q}$ as a function of the applied current. In all the panels we have used three values of the inhomogeneous interlattice exchange $A_{12} = 0, -5, -10\,\mathrm{pJ/m}$ . For the inhomogeneous intralattice

exchange, the values are $A_{11}=2$ pJ/m for (a) and (b), $A_{11}=6$ pJ/m for (c) and (d), and $A_{11}=15$ pJ/m for (e) and (f). The homogeneous interlattice exchange is $A_0=-2$ pJ/m for (a), (c) and (e) and $A_0=-15$ pJ/m for (b), (d) and (f). Lines are calculated by numerically solving Eq. (11) while dots are from full micromagnetic simulations.

*Dynamic properties: DW velocity.* Figure 4 compares the DW velocity $\dot{q}$ computed by numerically solving Eq. (11) with the one obtained from full micromagnetic simulations for a wide range of the exchange parameters (see figure caption). The agreement between micromagnetic simulations and one-dimensional calculations is very good with slight differences at very high current density $J>=7\,\mathrm{TA/m^2}$.

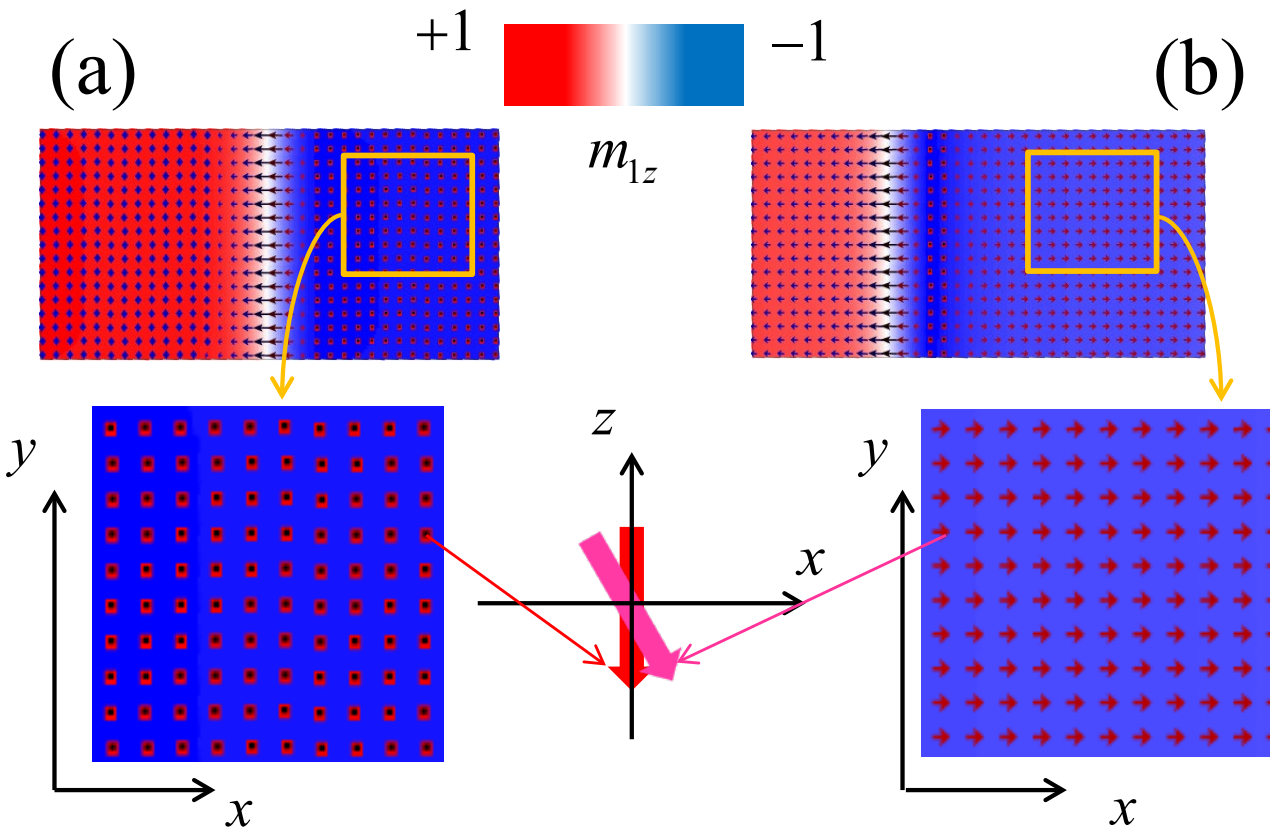

FIG. 5. Snapshots of the first sublattice magnetization from μM for (a) equilibrium state and (b) under a high current density ($7\,\mathrm{TA/m^2}$). In the latter, both domains acquire a non-negligible in-plane component affecting the reliability of the simplified models.

At such a high current density the domains themselves acquire a non-negligible in-plane component as can be seen in Fig. 5(a) and (b), so Eq. (8) is no longer valid. Differently from the FM case, here the linear behavior of the DW velocity is maintained at larger currents due to the stabilization role for the Néel configuration of the homogeneous exchange, analogously to the RKKY interaction in the case of synthetic antiferromagnets.[10] Even though the proposed model allows for misalignments of the in-plane components of the two sublattice magnetizations, no significant misalignments are observed for realistic parameters. Nevertheless, it is possible to get the condition for which this behavior is kept. To do that, we set the stationary conditions, $\dot{\Phi}_1=\dot{\Phi}_2=\dot{\Delta}/\Delta=0$. The sum of the dynamic equations for the in-plane angles give us the condition $\sin\Phi_1=\sin\Phi_2$. It can be checked that $\Phi_1=\Phi_2$ is unstable so $\Phi_1=\pi-\Phi_2$. Because of this relation the third equation from Eq. (11) can be rewritten as

$$\tan\Phi_1=\frac{\dfrac{\pi h_{SH}}{2\alpha\left(h_{exch}'+h_{exch}\right)}}{\dfrac{\pi}{2}h_D/\left(h_{exch}'+h_{exch}\right)+2\cos\Phi_1} \tag{16}$$

From Eq. (16), it is trivial to demonstrate that if $h_{SH} << \alpha \left| h_{exch}' + h_{exch} \right|$

$\Phi_1 = 0, \pi$, $\Phi_2 = \pi - \Phi_1$, the DW width remains the one at equilibrium (see Eq. (15)) and the linear behavior for the DW velocity is recovered (see Eq. (14)).

*Dynamic properties: DW nucleation.* We observe that it exists a maximum current density, $J_{th} = 8$ TA/m$^2$ for $A_{11} = 2$pJ/m ( $I \approx 160$ mA ), that can be applied without leading to the nucleation of other domains at the edges. Supplementary Movie 1[46] shows these dynamics achieved for $J = 9$ TA/m$^2$ (approximately $I \approx 180$ mA ). This DW nucleation from the edge driven by the current, already observed in FM,**¡Error! Marcador no definido.** is determined by the iDMI boundary conditions (in fact simulations without those boundary conditions show no DW nucleation, see Supplementary Movie 2[46]). In AFM, this mechanism is more efficient due to the stabilization of the $x$-component of the magnetization. In other words, in FM the magnetization at the edge rotates towards the $y$-direction reducing the SOT, but in AFM this rotation does not take place because of the antiferromagnetic exchange. A systematic study of the domain nucleation at the edge as a function of the iDMI parameter, $D$, for different intralattice and interlattice inhomogeneous exchange interactions, $A_{11}$ and $A_{12}$ respectively, is summarized in Fig. 6. The $x$-component of the sublattice magnetization in absolute value (the same for both) is displayed in Fig. 6(a). It increases as a function of $D$ and, on the other hand, decreases as a function of $A_{11}$ and $|A_{12}|$. This tilting originates a torque at the edge (roughly proportional to the tilting angle) promoting the nucleation of the domain with the opposite sign of $z$. As a consequence, the minimum threshold current density $J_{TH}$ for domain nucleation decreases as a function of $D$ (see data summarized in Fig. 6(b)).

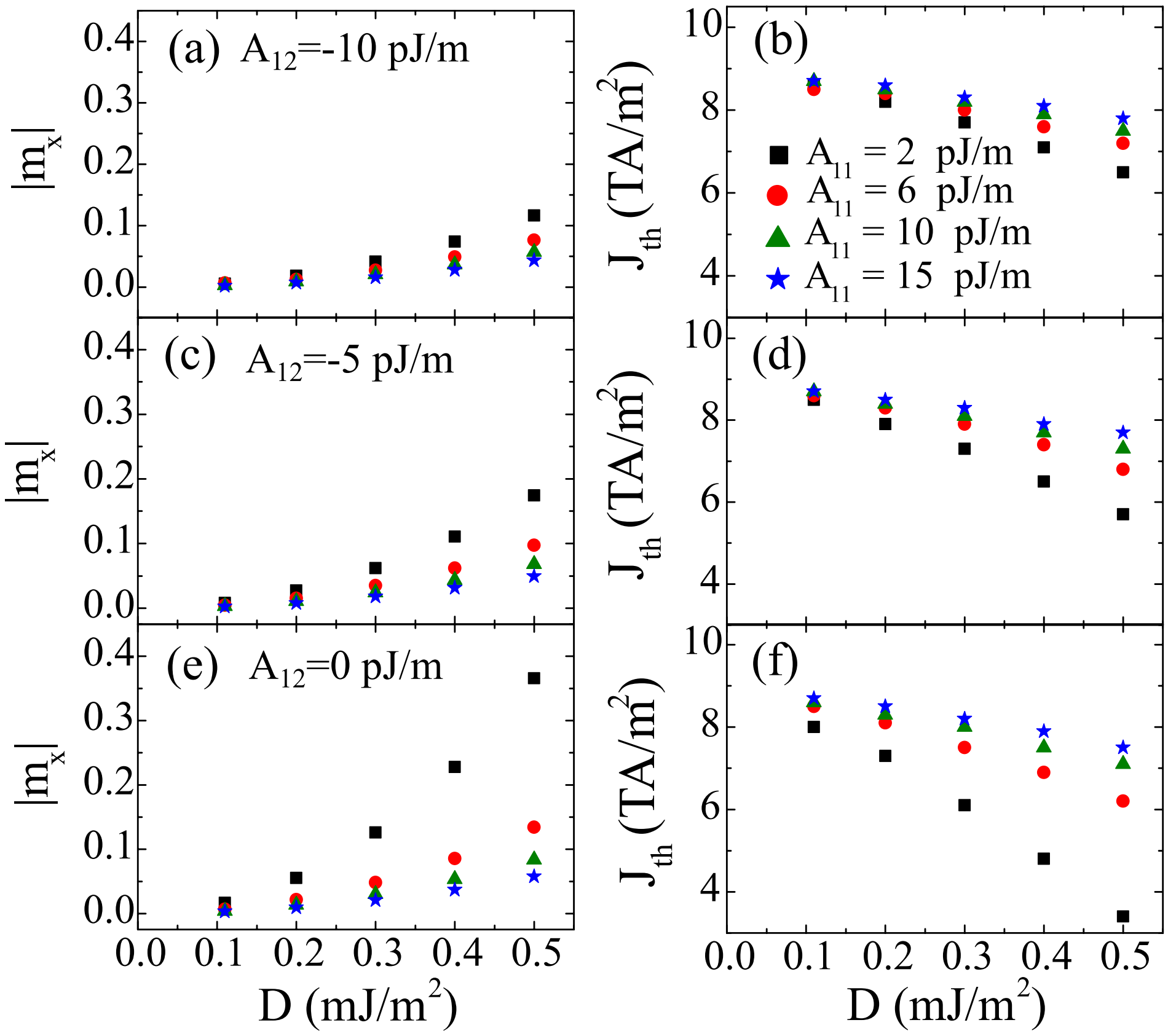

FIG. 6. (a), (c) and (e) $x$-component of the sublattice magnetization, responsible of the nucleation of domains at the edge. (b), (d) and (f) threshold current density needed to nucleate new domains as a function of the iDMI parameter, $D$, for different inhomogeneous intralattice parameters, $A_{11}$. Bottom panels stands for a inhomogeneous interlattice parameters $A_{12}=0$ pJ/m, middle panels for $A_{12}=-5$ pJ/m, and top panels for $A_{12}=-10$ pJ/m.

As the current increases, the DWs also acquire a slight curvature (see Supplementary Movie 1[46]) due to the smaller torques at the edges caused by the reduction of the $x$-component and the increase of the $y$-component of the magnetization.

We conclude that, in antiferromagnetic racetrack memories, the domain nucleation from the edges is the mechanism limiting the maximum velocity of an AFM DW, at least without changing the numbers of DWs and hence the information content of the racetrack itself.

*Dynamic properties: Role of exchange contributions.* In this section, we show the results of a systematic study of the DW velocity as a function of the different exchange interactions. Figures 7(a)-(c) summarize a comparison (full micromagnetic and the generalized simplified model) for a current density $J=1\,\text{TA/m}^2$, and a good agreement is observed in a wide range of parameters. The solid lines are from the generalized simplified model calculations while the dots indicate the full numerical computations. A main result is that the DW velocity is insensitive to the homogeneous exchange at low currents (Figure 7(a)), provided it is large enough to avoid misalignments between the magnetization of the two sublattices. On the other hand, the DW velocity is a square root function of both inhomogeneous terms, trend originated by the proportionality with the DW width $\Delta$ (yellow dashed lines in Fig. 7(b) and (c)), see Eq. (14). This demonstrates the inhomogeneous terms modify the DW width parameter without changing the DW structure in the stationary state determined by the two in-plane angles $\Phi_1$ and $\Phi_2$. Since the DW velocity is proportional to the DW width (see Eq. (14)), the induced increase of the DW width leads to a larger DW velocity. For larger currents, the DW velocity is expected to depend on the homogeneous exchange by the DW width dependence on this contribution (see Eq. (13) which predicts a decreasing trend when the two sublattice magnetizations align). Nevertheless, we are well below the condition $h_{SH} << \alpha\left(h_{exch}' + h_{exch}\right)$ and we only observe a small dependence for the largest currents considered in this work, $J_{th} = 7$ TA/m$^2$ (not show here).

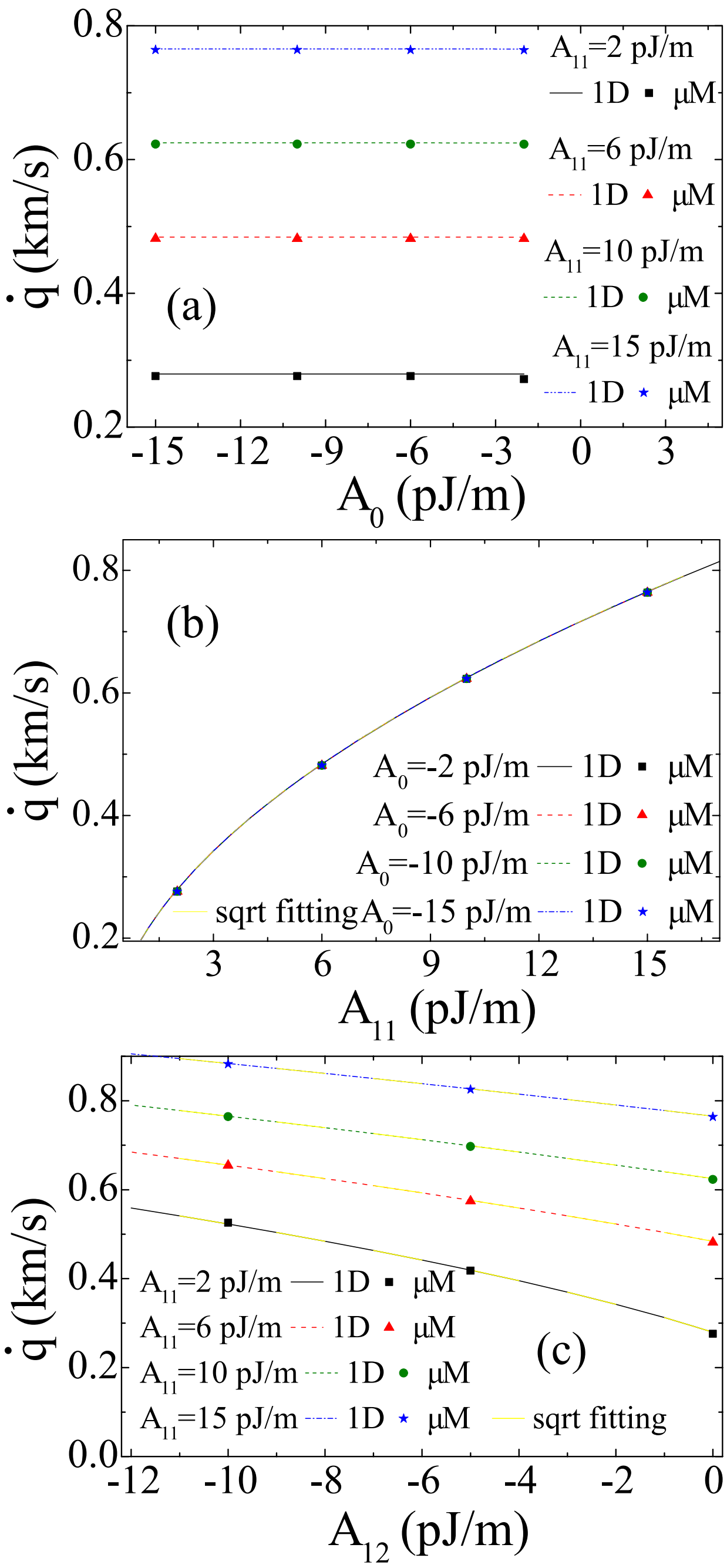

FIG. 7. DW velocity computed for a current density $J = 1\,\mathrm{TA/m^2}$ as a function of the exchange interactions, (a) dependence on the homogeneous interlattice coefficient for different values of $A_{11} = 2,6,10,15\,\mathrm{pJ/m}$, (b) dependence on the inhomogeneous intralattice coefficient for different values of $A_0 = -2,-6,-10,-15\,\mathrm{pJ/m}$ and (c) dependence on the inhomogeneous interlattice coefficient for different values of $A_{11} = 2,6,10,15\,\mathrm{pJ/m}$

*Dynamic properties. Interlattice damping.* Finally, it has been considered the role of an interlattice damping parameter $\alpha_{12}$, [47] which enters in the LLG of Eq. (1) in the following way

$$\begin{cases} \dot{\boldsymbol{m}}_1 = -\boldsymbol{m}_1 \times \boldsymbol{h}_{eff,1} + \boldsymbol{\tau}_1 + \alpha_{11}\boldsymbol{m}_1 \times \dot{\boldsymbol{m}}_1 + \alpha_{12}\boldsymbol{m}_1 \times \dot{\boldsymbol{m}}_2 \\ \dot{\boldsymbol{m}}_2 = -\boldsymbol{m}_2 \times \boldsymbol{h}_{eff,2} + \boldsymbol{\tau}_2 + \alpha_{22}\boldsymbol{m}_2 \times \dot{\boldsymbol{m}}_2 + \alpha_{21}\boldsymbol{m}_2 \times \dot{\boldsymbol{m}}_1 \end{cases} \tag{17}$$

where the intralattice damping parameter has been renamed for a clearer description ( $\alpha = \alpha_{11} = \alpha_{22}$ ). In spherical coordinates, Eq. (17) becomes:

$$\begin{cases} \dot{\theta}_i = -\dfrac{1}{L\sin\theta_i}\dfrac{\delta u}{\delta\varphi_i} - \alpha_{11}\sin\theta_i\dot{\varphi}_i - \alpha_{12}\cos\left(\varphi_i + \varphi_j\right)\sin\theta_i\dot{\varphi}_j + \dfrac{1}{L}h_{SH}\cos\theta_i\sin\varphi_i \\ \\ \sin\theta_i\dot{\varphi}_i = \dfrac{1}{L}\dfrac{\delta u}{\delta\theta_i} + \alpha_{11}\dot{\theta}_i + \dfrac{1}{L}h_{SH}\cos\varphi_i - \alpha_{12}\left(\sin^2\theta_i\cos\varphi_i\dot{\varphi}_j - \cos^2\theta_i\sin\varphi_i\dot{\theta}_j\right)\sin\varphi_j \end{cases} \tag{18}$$

Making the same assumption as in the previous case (Walker ansatz, $q_1 = q_2 = q$ , and $\Delta_1 = \Delta_2 = \Delta$ ). Eq. (11) is now

$$\begin{aligned} &\alpha_{11}\frac{\dot{q}}{\Delta} + \left(1 + \alpha_{12}\sin\Phi_2\cos\Phi_1\right)\frac{\dot{\Phi}_1}{2} - \left(1 + \alpha_{12}\sin\Phi_2\cos\Phi_1\right)\frac{\dot{\Phi}_2}{2} = \frac{\pi}{2}\frac{h_{SH}}{L_T}\left(\cos\Phi_1 - \cos\Phi_2\right) \\ &-L_T\left(\left(\frac{\pi^2}{6}\alpha_{11} - \frac{\alpha_{12}}{2}\sin\Phi_1\sin\Phi_2\right)\frac{\dot{\Delta}}{\Delta} + \frac{2}{3}\alpha_{12}\left(\sin\Phi_2\cos\Phi_1\dot{\Phi}_2 + \sin\Phi_1\cos\Phi_2\dot{\Phi}_1\right)\right) = \\ &\quad = 4K_u - \frac{4A_{11}}{\Delta^2} + \frac{2A_{12}}{\Delta^2}\left(\frac{2 - \cos\left(\Phi_1 - \Phi_2\right)}{3}\right) - 2h_{exch}\left(\cos\left(\Phi_1 - \Phi_2\right) + 1\right) + 2H_d' \\ &L_T\Delta\left[Q\frac{\dot{q}}{\Delta} - \alpha_{11}\dot{\Phi}_1 - \alpha_{12}\cos\left(\Phi_1 + \Phi_2\right)\dot{\Phi}_2\right] = -Q\pi D\sin\Phi_1 + 2h_{exch}\Delta\sin\Phi_d \\ &L_T\Delta\left[-Q\frac{\dot{q}}{\Delta} - \alpha_{11}\dot{\Phi}_2 - \alpha_{12}\cos\left(\Phi_1 + \Phi_2\right)\dot{\Phi}_1\right] = Q\pi D\sin\Phi_2 - 2h_{exch}\Delta\sin\Phi_d \end{aligned} \tag{19}$$

Because the considerations made previously to derive Eq. (16) are still valid, we can consider that $\Delta$ is constant and equal to the equilibrium value for low currents. Then we can omit the second equation. A fast exploration of the system including interlattice damping parameter (not shown here), with $\alpha_{12} = 0.01, 0.05, 0.09, 0.0999$ , shows no changes on the stationary values. Moreover, we observe that the interlattice damping only affects the terms which are zero at the stationary regime, so no changes are expected for the stationary DW velocity. Nevertheless, we observe small changes in the transient regime even for low currents so these new terms could be important when considering other conditions, such as AFM oscillators.[18,21,22]

## V. Conclusions

Velocities up to a few km/s for antiferromagnetic domain walls have been predicted making antiferromagnets a testbed material for the development of ultrafast racetrack memories and THz spintronic devices. Here, we have extended results of previous works on this topic, by deriving a generalized expression for DW width and velocity that has been benchmarked with continuous micromagnetic simulations in a wide range of parameters. A systematic study of the role of different exchange interactions shows a DW velocity independent of the homogeneous interlattice exchange at low currents, and with a square root dependence on both inhomogeneous exchanges, i.e. intralattice and interlattice. This dependence is inherited from the behavior of the DW width, which is predicted to decrease at high currents due to the homogeneous interlattice exchange. Finally, we show that the domain wall velocity in an antiferromagnetic racetrack memory will be limited by the nucleation of new domains at the edges of the system, due to the iDMI boundary conditions that, for example in racetrack memories, can change the content of stored information. Therefore, it should be noticed that even a small iDMI parameter is needed to promote the Néel type wall, large $D$ values are undesirable as they would lead to lower threshold currents. On the contrary, high inhomogeneous exchange interaction would increase the threshold current, and also the DW width, promoting higher DW velocities. Nevertheless, the larger DW width would increase the minimum domain size and then decrease the storage density. The analytical approach employed here can be used as a starting point for the development of a one-dimensional model for the description of DW motion in ferrimagnets.

## Acknowledgements

G.F. and M.C. would like to acknowledge the contribution of the COST Action CA17123 "Ultrafast opto-magneto-electronics for non-dissipative information technology". P.K.A. acknowledges support by a grant from the U.S. National Science Foundation, Division of Electrical, Communications and Cyber Systems (NSF ECCS-1853879). This work was also supported by PETASPIN. The authors also acknowledge Nick Kioussis for the discussion regarding the interlattice damping.

## APPENDIX A: SIGMA MODEL

The DW width $\Delta$, the DW position $q$, and the in-plane angle of the magnetization of the Néel order parameter, $\Phi$ for the sigma model are defined in Fig. 2(a) and (b). Equation (1) can be rewritten in terms of the Néel order parameter $\boldsymbol{l}=\boldsymbol{m}_1-\boldsymbol{m}_2$ and the small magnetization $\boldsymbol{m}=\boldsymbol{m}_1+\boldsymbol{m}_2$ as:[13,35,42]

$$\dot{\boldsymbol{m}}=-\gamma_0\left(\boldsymbol{m}\times\boldsymbol{H}_m+\boldsymbol{l}\times\boldsymbol{H}_l\right)+\frac{\alpha}{2}\left(\boldsymbol{m}\times\dot{\boldsymbol{m}}+\boldsymbol{l}\times\dot{\boldsymbol{l}}\right)-\frac{\gamma_0 H_{SH}}{2}\left(\boldsymbol{m}\times\left(\boldsymbol{m}\times\boldsymbol{p}\right)+\boldsymbol{l}\times\left(\boldsymbol{l}\times\boldsymbol{p}\right)\right) \quad \text{(A1.a)}$$

$$\dot{\boldsymbol{l}}=-\gamma_0\left(\boldsymbol{l}\times\boldsymbol{H}_m+\boldsymbol{m}\times\boldsymbol{H}_l\right)+\frac{\alpha}{2}\left(\boldsymbol{l}\times\dot{\boldsymbol{m}}+\boldsymbol{m}\times\dot{\boldsymbol{l}}\right)-\frac{\gamma_0 H_{SH}}{2}\left(\boldsymbol{l}\times\left(\boldsymbol{m}\times\boldsymbol{p}\right)+\boldsymbol{m}\times\left(\boldsymbol{l}\times\boldsymbol{p}\right)\right) \quad \text{(A1.b)}$$

where the dot convention for the time derivative has been adopted and $\boldsymbol{H}_m$ and $\boldsymbol{H}_l$ are the effective fields with respect to $\boldsymbol{m}$ and $\boldsymbol{l}$.

Let's start with a simplified formulation, where the energy density $u$ has the following expression:

$$u = A_{11}\left(\nabla \boldsymbol{l}\right)^2 - \frac{A_0}{a^2}\boldsymbol{m}^2 - \frac{K_u}{2} l_z^2 + \frac{D}{2}\left(m_z\left(\nabla\cdot\boldsymbol{m}\right) - \left(\boldsymbol{m}\cdot\nabla\right)m_z + l_z\left(\nabla\cdot\boldsymbol{l}\right) - \left(\boldsymbol{l}\cdot\nabla\right)l_z\right). \qquad \text{(A2)}$$

The expression derived in Ref. 38 is obtained neglecting the $\boldsymbol{l}$-dependence of the homogeneous exchange and $\boldsymbol{m}$-dependence of the uniaxial anisotropy have been neglected and assuming that $-A_{12} = 2A_{11}$. From Eq. (A1.b) it is possible to determine $\boldsymbol{m}$ as a function of $\boldsymbol{l}$ considering the anisotropy term and the spatial derivatives much smaller than the other terms,[38] neglecting dissipative terms,[35] and by taking into account that $\boldsymbol{l}\times\left(\boldsymbol{m}\times\boldsymbol{l}\right) = \boldsymbol{m}\,\boldsymbol{l}^2 \approx 4\boldsymbol{m}$. Inserting this expression in Eq. (A1.a), the dynamics of $\boldsymbol{l}$ does not depend on $\boldsymbol{m}$. Thus, Eq. (A1.a) in spherical coordinates for $\boldsymbol{l}$ reads

$$\begin{cases} \ddot{\theta} - c^2\theta'' + \sin\theta\cos\theta\left(c^2\varphi'' - \dot{\varphi}^2\right) + b^2\sin\theta\cos\theta + d^2\sin^2\theta\sin\varphi\,\varphi' = \\ = +2\gamma_0\alpha H_{exch}\dot{\theta} + 2\gamma_0^2 H_{SH} H_{exch}\cos\varphi \\ \\ \dfrac{d}{dt}\left(\sin^2\theta\dot{\varphi}\right) - c^2\dfrac{d}{dx}\left(\sin^2\theta\varphi'\right) - d^2\sin^2\theta\sin\varphi\,\theta' = \\ = -2\gamma_0\alpha H_{exch}\sin^2\theta\,\dot{\varphi} + 2\gamma_0^2 H_{SH} H_{exch}\sin\theta\cos\theta\sin\varphi \end{cases} \qquad \text{(A3)}$$

with $c^2 = -\left(\dfrac{2\gamma_0}{\mu_0 M_s}\right)^2 2\dfrac{A_0}{a^2}2A_{11}$ , $b^2 = -\left(\dfrac{2\gamma_0}{\mu_0 M_s}\right)^2 2\dfrac{A_0}{a^2}K_u$ , $d^2 = -\left(\dfrac{2\gamma_0}{\mu_0 M_s}\right)^2 2\dfrac{A_0}{a^2}D$ , $H_{exch} = \dfrac{2A_0}{a^2\mu_0 M_s}$ and ' standing for $x$ partial derivative. At equilibrium, one exact solution of the system of equations (A3) is the Walker ansatz,[48] which describes an approximate DW profile:

$$\begin{aligned} \theta(t,x) &= 2\arctan\left(\exp\left(Q\frac{x-q(t)}{\Delta}\right)\right) \\ \Phi(t,x) &= \Phi(t) \end{aligned} \qquad \text{(A4)}$$

where $q$, and $\Delta$ are again the DW position and width and now $\Phi$ stands for the in-plane angle of the Néel order parameter, as have been defined in Figs. 2(a) and (b). $Q = \pm 1$ allows distinguishing between an up-down transition ($Q = 1$) or a down-up transition ($Q = -1$). Under the hypothesis that Eq. (A4) is still valid for moving DWs, it is possible to derive a couple of equations for $q$ and $\Phi$ which, at stationary conditions ( $\ddot{q} = \ddot{\Phi} = \dot{\Phi} = 0$), transforms into:

$$\begin{aligned} \dot{q} &= Q\frac{\pi}{2}\frac{\gamma_0\Delta H_{SH}}{\alpha}\cos\Phi \\ H_D\sin\Phi &= 0 \Rightarrow \Phi = 0,\pi \end{aligned} \qquad \text{(A5)}$$

where $H_D = D/\mu_0 M_S$. The actual solution for $\Phi$ ( 0 or $\pi$ ) is determined by the sign of the iDMI and the sign of $Q$, while the modulus of the DW velocity is then:

$$|\dot{q}| = \frac{\pi}{2}\frac{\gamma_0 \Delta |H_{SH}|}{\alpha} \quad \text{(A6)}$$

We stress that this equation is valid within the assumptions previously made, which are fulfilled for large enough $A_0$ to maintain the $\Phi_1$ and $\Phi_2$ at 0 and $\pi$ for any applied $J$.

## References

[1] S. S. P. Parkin, M. Hayashi, and L. Thomas, *Science* 320, 190 (2008).
[2] E. Martinez, S. Emori, N. Perez, L. Torres, and G. S. D. Beach. *J. Appl. Phys,* 115: 213909, (2014).
[3] X. Wang, Y. Chen, H. Xi, H. Li, and D. Dimitrov, IEEE Elect. Dev. L, 30, 3, (2009)
[4] S. Lequeux, J. Sampaio, V. Cros, K. Yakushiji, A. Fukushima, R. Matsumoto, H. Kubota, S. Yuasa and J. Grollier, *Sci. Rep.* 6, 31510 (2016).
[5] J. Cai, B. Fang, C. Wang, and Z. Zeng, *Appl. Phys. Lett*. 111, 182410 (2017)
[6] A. Cao, X. Zhang, Z. Li, Q. Leng, L. Wen, and W. Zhao, *IEEE Magn. Lett.,* 9, 1404804 (2018).
[7] N. L. Schryer and L. R. Walker, *J. Appl. Phys*., 45(12):5406, (1974).
[8] A. Thiaville, S. Rohart, E. Jue, V. Cros, and A. Fert. *EPL* 100(5):57002, (2012).
[9] S.H. Yang, K.S. Ryu, and S. Parkin. *Nat. Nanotechnol.*, 10:221-226, (2015).
[10] R Tomasello, V. Puliafito, E. Martinez, A. Manchon, M. Ricci, M Carpentieri and G. Finocchio *J. Phys. D*, 50(32):325302, (2017).
[11] S. A. Siddiqui, J. Han, J. T. Finley, C. A. Ross, and L. Liu, *Phys. Rev. Lett*., 121, 057701 (2018).
[12] L. Caretta, M. Mann, F. Büttner, K. Ueda, B. Pfau, C. M. Günther, P. Hessing, A. Churikova, C. Klose, M. Schneider, *et. al*., *Nat. Nanotechnol* ., 13, 1154–1160 (2018).
[13] O. Gomonay, M. Kläui, and J. Sinova, *Appl. Phys. Lett*,109, 142404 (2016).
[14] T. Shiino, S-H Oh, P. M. Haney, S-W Lee, G. Go, B-G Park, and K-J Lee, *Phys. Rev. Lett*., 117, 087203 (2016).
[15] O. Gomonay, T. Jungwirth, and J. Sinova, *Phys. Rev. Lett*, 117, 017202 (2016).
[16] C. Ó. Coileáin and H. C. Wu, SPIN, 7, 3 1740014 (2017).
[17] A. H. Macdonald, and M. Tsoi, *Philos. Trans. R. Soc. A* 369, 3098 (2011).
[18] E. V. Gomonay and V. M. Loktev, *Low Temp. Phys*. 40, 17 (2014).
[19] T. Jungwirth, X. Marti, P. Wadley. and J. Wunderlich, *Nat. Nanotechnol*.11, 231 (2016).
[20] T. Jungwirth, J. Sinova, A. Manchon, X. Marti, J. Wunderlich, and C. Felser, *Nat. Phys*. 14, 200-203 (2018).
[21] R. Khymyn, I. Lisenkov, V. Tiberkevich, B. A. Ivanov and A. Slavin, *Sci. Rep*. 7, 43705 (2017).
[22] V. Puliafito, R. Khymyn, M. Carpentieri, B. Azzerboni, V. Tiberkevich, A. Slavin, and G. Finocchio *Phys. Rev. B* 99, 024405 (2019).
[23] A. V. Kimel, B. A. Ivanov, R. V. Pisarev, P. A. Usachev, A. Kirilyuk and Th. Rasing, Nat. Phys. 5, 727–731 (2009).
[24] P. Wadley, B. Howells, J. Železný, C. Andrews, V. Hills, R. P. Campion, V. Novák, K. Olejník, F. Maccherozzi, S. S. Dhesi, et al., *Science* 10.1126/science.aab1031 (2016).
[25] S.Yu. Bodnar, L. Šmejkal, I. Turek, T. Jungwirth, O. Gomonay, J. Sinova, A.A. Sapozhnik, H.-J. Elmers, M. Kläui and M. Jourdan. *Nat. Commun.* , 9, 348 (2018).
[26] C. Wang, H. Seinige, G. Cao, J.-S. Zhou, J. B. Goodenough, and M. Tsoi, *Phys. Rev. X*, 4, 041034 (2014).
[27] J. Godinho, H. Reichlová, D. Kriegner, V. Novák, K. Olejník, Z. Kašpar, Z. Šobáň, P. Wadley, R. P. Campion, R. M. Otxoa,, *et. al., Nat. Commun*., 9, 4686 (2018).
[28] T. Moriyama, K. Oda, T. Ohkochi, M. Kimata and T. Ono, *Sci. Rep.*, 8:14167 (2018).
[29] D. Suess, T. Schrefl, W. Scholz, J.-V. Kim, R. L. Stamps, and J. Fidler, IEEE Trans. Magn. 38(5) 2397-2399 (2002).
[30] B. Fang *et. al.*, *Phys. Rev. Applied* 11, 014022 (2019).

[31] N. Ntallis, K.G. Efthimiadis, *Computational Materials Science*, 97, 42 (2015).
[32] L. Liu, T. Moriyama, D. C. Ralph, and R. A. Buhrman, Phys. Rev. Lett. 106, 036601 (2011).
[33] J. E. Hirsch. *Phys. Rev. Lett.*, 83:1834-1837, (1999).
[34] N. Perez, L. Torres, and E. Martinez-Vecino. *IEEE Trans. Magn.,* 50(11):1-4, (2014).
[35] H. V. Gomonay and V. M. Loktev, *Phys. Rev. B* 81, 144427 (2010).
[36] S. Emori, U. Bauer, S.M. Ahn, E. Martinez, and G. S. D. Beach. *Nat. Mater.*, 12:611, (2013).
[37] A. Manchon, I.M. Miron, T. Jungwirth, J. Sinova, J. Zelezny, A. Thiaville, K. Garello, and P. Gambardella, arXiv:1801.09636 (2018).
[38] A. Kosevich, B. Ivanov, and A. Kovalev*, Phys. Rep.* 194, 117 (1990).
[39] E. Martinez, S. Emori, and G. S. D. Beach. *Appl. Phys. Lett*, 103:072406, (2013).
[40] L. Lopez-Diaz, D. Aurelio, L. Torres, E. Martinez, M. A. Hernandez-Lopez, J. Gomez, O. Alejos, M. Carpentieri, G. Finocchio and G. Consolo., J. Phys. D: Appl. Phys. 45, 323001 (2012).
[41] J. Miltat and M. J. Donahue, Handbook of Magnetism and Advanced Magnetic Materials (Wiley, New York, 2007), Vol. 2, pp. 742–764.
[42] E. G. Tveten, A. Qaiumzadeh, O. A. Tretiakov, and A. Brataas, *Phys. Rev. Lett*., 110, 127208 (2013).
[43] E. Martinez, L. Torres, N. Perez, M. A. Hernandez, V. Raposo, and S. Moretti. *Sci. Rep.*, 5:10156, 6 (2015).
[44] S. Rohart and A. Thiaville. *Phys. Rev. B,* 88:184422, (2013).
[45] A. Thiaville, J. M. García, and J. Miltat. *J. Magn. Magn. Mater*., 242-245, 1061-1063, (2002).
[46] See Supplemental Material at
[47] F. Mahfouzi and N. Kioussis, *Phys. Rev. B,* 98, 220410(R) (2018).
[48] E. G. Tveten, T. Müller, J. Linder and A. Brataas, *Phys. Rev. B,* 93, 104408 (2016).